# Anisotropy of the acoustooptic figure of merit for LiNbO$_3$ crystals. Isotropic diffraction

**Oksana Mys, Myroslav Kostyrko, Oleh Krupych and Rostyslav Vlokh***

*Vlokh Institute of Physical Optics, 23 Dragomanov Street, 79005 Lviv, Ukraine*
*\* Corresponding author: vlokh@ifo.lviv.ua*

We have developed the approach for analyzing anisotropy of acoustooptic figure of merit (AOFM) for lithium niobate crystals in the case of isotropic acoustooptic (AO) diffraction. The working relations for the effective elastooptic coefficients and the AOFM have been derived. We have found that, under the conditions of isotropic AO diffraction, the maximum AOFM value for LiNbO$_3$ is equal to $11.62 \times 10^{-15}\,\mathrm{s}^3/\mathrm{kg}$. It is peculiar for the geometry of AO interaction of the shear acoustic wave propagating in the *YZ* plane (the velocity 3994 m/s) with the optical wave polarized in the same plane. We have demonstrated that the maximum AOFM values are achieved mainly due to essential anisotropy and high values of the elastooptic coefficient of LiNbO$_3$.

## 1. Introduction

LiNbO$_3$ represents a crystalline material which is widely used in optoelectronics [1]. A well developed crystal growth technology and nearly perfect optical characteristics have made lithium niobate one of the most requested materials for various electrooptic [2, 3], nonlinear optical [3] and integrated optical [4] applications. The LiNbO$_3$ crystals are also a well known crystalline material for acoustooptic (AO) control of laser radiation [5]. It is interesting that exploitation of lithium niobate in AO cells is predetermined by its well-developed technology of obtaining and processing [1,6] rather than high AO figure of merit (AOFM). Good piezoelectric properties needed for efficient excitation of surface acoustic waves (AWs) represent another useful feature of this material, which is widely applied in acoustooptics and integrated optics [7]. Despite of the fact that AOFM for LiNbO$_3$ crystals are not high enough, these crystals exhibit other properties that make them promising materials for AO application. Lithium niobate crystals possess low acoustic attenuation in the high acoustic wave frequency range [5,8], which leads to the possibility of utilizing of these crystals in particular in the integrated optical waveguide AO devices [7]. For example at the acoustic wave frequency equal to 1 GHz the acoustic attenuation is about 1 dB/cm [8]. From other side these crystals can be used for wide band AO interaction with utilization of different techniques, i.e. multiple surface acoustic wave (SAW) transducers, multiple tilted SAW transducers, phased SAW's, stepped SAW's, etc [8, 9–11].

The highest AOFM for the LiNbO$_3$ crystals ($2.92 \times 10^{-15}$ s$^3$/kg) has been achieved for the case of AO interaction with the transverse AW that propagates along the crystallographic axis $c$ [12] with the velocity $v_{32} = 3574$ m/s. Here the first and the second indices correspond respectively to the AW propagation and polarization directions. Notice that the AW mentioned above is indeed among the slowest waves that can propagate in LiNbO$_3$. Nonetheless, the corresponding AOFM, which is proportional to the cube of AW slowness ($M_2 \propto v_{ij}^{-3}$), is not high enough when compared to the parameters typical for the best AO materials (e.g., for TeO$_2$ crystals manifesting the highest AOFM equal to ~$1200 \times 10^{-15}$ s$^3$/kg [13]).

As a result, optimization of AO interaction geometry and improving the AOFM parameter are necessary in order to achieve high AO efficiencies at low levels of driving signals. We note that some results in this direction have already been reported in the literature. For example, the attempts at the analyses of AOFM anisotropy for the LiNbO$_3$ crystals have been made in the works [14, 15]. However, the technique suggested in those studies is not transparent enough, since it is not clear how to use it in practice, the results of Refs. [14, 15] are based on the incorrect assumption that one can determine the AOFM anisotropy by merely rewriting the elastooptic tensor in some rotated coordinate system. In addition, authors [14, 15] have considered only wave vectors but not all tensors in the spherical coordinate system; the Bragg condition is not satisfied, in particular for those interactions which lead to the extreme values of AOFM; formula for AOFM which take into account the acoustic and optical beams oblique propagation has not relation to this effect (see [11]). Finally, the authors of the mentioned works have not presented the relations for the effective elastooptic coefficients (EECs). These makes the papers [14,15] completely wrong and inappropriate for further discussion. Interesting and reliable results for the AOFM anisotropy in lith-ium niobate have been obtained in Ref. [16], although the authors have analyzed only the anisotropic AO interactions with the longitudinal AWs.

In our recent works [17–19] we have developed a method for comprehensive analysis of AOFM anisotropy and have approved it on the example of TeO$_2$ crystals [18, 19]. As a matter of fact, those calculation results agree well with the experimental data. In the present study we report the results of analysis of the AOFM anisotropy performed for the case of trigonal crystals that belong to the point groups of symmetry 3m, 32 and $\overline{3}$m on the example of LiNbO$_3$ crystals.

## 2. Methods of analysis

Before proceeding with explanations of our analytical method, we remind in brief the main properties of LiNbO$_3$ needed for our further analysis. Lithium niobate belongs to the point symmetry group 3m. One of its mirror symmetry planes is perpendicular to the crystallographic axis $a$ [20] and the axis $c$ is parallel to the three-fold symmetry axis. Further on, we accept that the crystallographic axes $a$, $b$ and $c$ corresponds respectively to the axes $X$, $Y$ and $Z$ of the optical Fresnel ellipsoid. LiNbO$_3$ is optically negative, with the refractive indices amounting to $n_o = 2.286$ and $n_e = 2.203$ ( $\lambda = 632.8$ nm ) [20]. The elastooptic coefficients determined in our recent work [21] at $\lambda = 632.8$ nm are equal to $p_{11} = -0.023 \pm 0.017$, $p_{12} = 0.076 \pm 0.014$, $p_{13} = 0.147 \pm 0.019$, $p_{31} = 0.157 \pm 0.007$, $p_{33} = 0.141 \pm 0.013$, $p_{14} = 0.057 \pm 0.004$, $p_{41} = 0.051 \pm 0.011$, and $p_{44} = 0.126 \pm 0.004$. We have to notice, that as it was shown in our work [21], the contribution of the secondary elastooptic effect to the elastooptic coefficients in the lithium niobate crystals is comparable with the experimental error of their determination. The elastic stiffness coefficients at the constant electric field are as follows: $C_{11} = 2.03$, $C_{12} = 0.573$, $C_{13} = 0.752$, $C_{33} = 2.42$, $C_{44} = 0.595$, $C_{66} = 0.728$, and $C_{14} = 0.085$ (all in the units of $10^{11}$ N/m$^2$). It is known that only two stiffness coefficients in LiNbO$_3$ crystals, namely $C_{33}$ and $C_{13}$, undergo the contribution of piezoelectricity at the calculation of elastic modules on the basis of certain set of acoustic wave velocities [22]. We have to notice that error of determination of these coefficients (i.e. $C_{13}$ and $C_{33}$) according to work [22] exceeds the difference between values of coefficients taken by us from [23] and their values presented in above mentioned paper. This was the reason, why we have used the stiffness coefficients determined at the conditions of the constant electric field for calculation of the acoustic wave velocities. Finally, the density of the crystal is equal to $\rho = 4640 \text{ kg/m}^3$ [23].

It is well known that the AOFM is given by the relation

$$M_2 = \frac{n^6 p_{ef}^2}{\rho v_{ij}^3},\tag{1}$$

where $p_{ef}$ is the EEC. It is readily seen from Eq. (1) that the main contribution to the AOFM should come from the AW velocity and the EEC. In principle, optimal values of these parameters can provide an interaction geometry for which the highest $M_2$ value is reached. In particular, the highest AOFM can be typical for the interactions with the slowest AW [24].

All the components of Christoffel matrices for the AW propagation in the *XZ* and *XY* planes in lithium niobate are non-zero, and so their eigenvalues cannot be obtained analytically in general. The analytical relations between the AW velocities and the elastic stiffness coefficients can be obtained only for the AW propagation in the *YZ* plane, which can be transferred to an arbitrary interaction plane by rotating the *YZ* plane around the *Z* or *Y* axes, with a following numerical determination of the AW velocities.

The dependences of the quasi-transverse AW velocities on the wave vector orientation in the *YZ* plane are given by the formulas

$$v_{QT_1}^2(\theta_Y) = \frac{(C_{44}+C_{11})\cos^2\theta_Y + \sin^2\theta_Y(C_{44}+C_{33}) - C_{14}\sin 2\theta_Y}{2\rho}$$
$$-\frac{\sqrt{[(C_{44}-C_{11})\cos^2\theta_Y + (C_{44}-C_{33})\sin^2\theta_Y + C_{14}\sin 2\theta_Y]^2 + 4[\cos\theta_Y\sin\theta_Y(C_{13}+C_{44}) + C_{14}\cos^2\theta_Y]^2}}{2\rho},\tag{2}$$

$$v_{QT_2}^2(\theta_Y) = \frac{C_{44}\sin^2\theta_Y + C_{66}\cos^2\theta_Y + C_{14}\sin 2\theta_Y}{\rho}.\tag{3}$$

The same relation for the quasi-longitudinal AW velocities takes the form

$$v_{QL}^2(\theta_Y) = \frac{(C_{44}+C_{11})\cos^2\theta_Y + \sin^2\theta_Y(C_{44}+C_{33}) - C_{14}\sin 2\theta_Y}{2\rho}$$
$$+\frac{\sqrt{[(C_{44}-C_{11})\cos^2\theta_Y + (C_{44}-C_{33})\sin^2\theta_Y + C_{14}\sin 2\theta_Y]^2 + 4[\cos\theta_Y\sin\theta_Y(C_{13}+C_{44}) + C_{14}\cos^2\theta_Y]^2}}{2\rho},$$
$$\tag{4}$$

where $\theta_Y$ denotes the angle between the AW vector and the *Y* axis in the *YZ* plane (i.e., the angle of rotation of the AW vector around the *X* axis – see Fig. 1a).

The other interaction planes can be regarded as being rotated by some angles $\varphi_Z$ around the $Z$ axis or by $\varphi_X$ around the $X$ axis (see Fig. 1b, c). In the new coordinate systems $X'Y'Z'$, the structure of the elastic stiffness tensor changes. The new tensor components can be derived after this tensor is rewritten in the new coordinate system according to a known procedure described in Ref. [25].

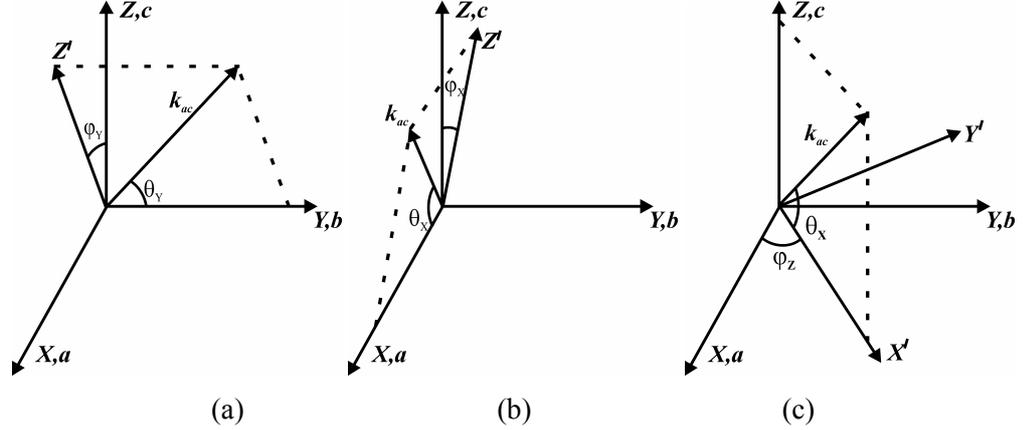

(a)         (b)         (c)

Fig. 1. A crystallographic coordinate system coupled with a Cartesian system $XYZ$, and new coordinate systems $X'Y'Z'$ obtained after rotating the interaction plane $YZ$ around the $Y$ axis (a), the $XZ$ plane around the $X$ axis (b), and the $XZ$ plane around the $Z$ axis (c).

As shown in our recent studies [17–19], nine types of the AO interactions can be implemented in crystals, six of them for the isotropic interactions and three for the anisotropic ones. In general, the six isotropic interactions can be presented as the interactions of the two optical eigenwaves having orthogonal polarizations with the three acoustic eigenwaves (two of them being transverse and one longitudinal). As an example, the six isotropic types of AO interactions occurring in the $XZ$ plane are as follows:

(*I*) the interaction of the longitudinal AW $v_{11} = v_{QL}$ propagating along the $X$ ($Y$) axis with the incident optical wave, of which electric induction vector is parallel to the $Y$ ($X$) axis;

(*II*) the interaction of the longitudinal AW $v_{11} = v_{QL}$ propagating along the $X$ ($Y$) axis with the incident optical wave, of which electric induction lies in the $XZ$ ($YZ$) plane at the Bragg angle $\theta_B$ with respect to the $X$ ($Y$) axis ( $D_3 = D\sin\theta_B$ and $D_1(D_2) = D\cos\theta_B$ );

(*III*) the interaction of the transverse AW $v_{13} = v_{QT_1}$ ($v_{23}$) propagating along the $X$ ($Y$) axis and polarized along the $Z$ axis with the incident optical wave polarized parallel to the $Y$

(*X*) axis (in present work we have replaced the types of interaction (*III*) and (*IV*) in respect to those, presented in [17, 18], since it seem to be more logical);

(*IV*) the interaction of the transverse AW $v_{13} = v_{QT_1}$ ($v_{23}$) propagating along the *X* (*Y*) axis and polarized along the *Z* axis with the incident optical wave, of which polarization vector lies in the *XZ* (*YZ*) plane at the angle $\theta_B$ with respect to the *X* (*Y*) axis ( $D_3 = D \sin \theta_B$ and $D_1(D_2) = D \cos \theta_B$ ;

(*V*) the interaction of the transverse AW $v_{12} = v_{QT_2}$ ($v_{21}$) propagating along the *X* (*Y*) axis and polarized along the *Y* (*X*) axis with the incident optical wave, of which polarization vector lies in the *XZ* (*YZ*) plane at the angle $\theta_B$ with respect to the *X* axis ( $D_3 = D \sin \theta_B$ and $D_1(D_2) = D \cos \theta_B$ ;

(*VI*) the interaction of the transverse AW $v_{12} = v_{QT_2}$ ($v_{21}$) propagating along the *X* (*Y*) axis and polarized along the *Y* (*X*) axis with the incident optical polarized parallel to the *Y* (*X*) axis.

The remaining three types of the anisotropic interactions in the *XZ* (or *YZ*) interaction plane are as follows:

(*VII*) the interaction of linearly polarized incident optical wave with the longitudinal AW, QL;

(*VIII*) the interaction of linearly polarized incident optical wave with the transverse AW, QT$_1$;

(*IX*) the interaction of linearly polarized incident optical wave with the transverse AW, QT$_2$.

Those anisotropic types of AO interactions are outside the scope of this work. They will be considered in detail in our forthcoming study.

## 3. Results and discussion

## 3.1. Effective elastooptic coefficients

Let us derive the analytical relations describing the EEC for different types of AO interactions in the LiNbO$_3$ crystals. These relations will be the same for all the crystals belonging to the point symmetry groups 3m, 32 and $\overline{3}$m . In order to consider the AO interactions in all of the interaction planes, the diffraction processes in the crystals belonging to the trigonal sys-tem have been considered in the interaction planes rotated around the $Z$ and $X$ axes by the angles $\varphi_Z$ and $\varphi_X$. However the additional rotation of the $YZ$ interaction plane around the $Y$ axis by the angle $\varphi_Y$ (see Fig. 1a) is needed in comparison with [18], since the axes $X$ and $Y$ are not equivalent from the viewpoints of acoustic and elastooptic properties. In particular, this situation refers to the point group 3m, i.e. the case of LiNbO$_3$.

For the interaction of the type (*I*), the components of the mechanical strain tensor caused by the AW QL that propagates initially along the $X$ axis in the $XZ$ plane can be written as

$$e'_1 = e_1 \cos^2\theta_X, \ e'_2 = e_1 \sin^2\theta_X, \ e'_5 = -e_1 0.5\sin 2\theta_X. \tag{5}$$

The corresponding EEC is given by $p_{ef}^{(I)} = p_{21}\cos^2\theta_X + p_{23}\sin^2\theta_X$. Here $\theta_X$ denotes the angle between the AW vector and the $X$ (or $X'$) axis. Due to rotation of the interaction plane around the $Z$ axis the components of the strain tensor become as follows:

$$\begin{aligned}
&e'_1 = e_1\cos^2\theta_X\cos^2\varphi_Z, \ e'_2 = e_1\cos^2\theta_X\sin^2\varphi_Z, \ e'_3 = e_1\sin^2\theta_X, \\
&e'_4 = -0.5e_1\sin 2\theta_X\sin\varphi_Z, \ e'_5 = -0.5e_1\sin 2\theta_X\cos\varphi_Z, \ e'_6 = 0.5e_1\cos^2\theta_X\sin 2\varphi_Z.
\end{aligned} \tag{6}$$

Then the EEC reads as

$$\begin{aligned}
p_{ef}^{(I)} = &\sin^2\varphi_Z\left(p_{11}\cos^2\theta_X\cos^2\varphi_Z + p_{12}\cos^2\theta_X\sin^2\varphi_Z + p_{13}\sin^2\theta_X + p_{14}\sin 2\theta_X\sin\varphi_Z\right) \\
&+ \cos^2\varphi_Z\left(p_{21}\cos^2\theta_X\cos^2\varphi_Z + p_{22}\cos^2\theta_X\sin^2\varphi_Z + p_{23}\sin^2\theta_X - p_{14}\sin 2\theta_X\sin\varphi_Z\right) \\
&+ \sin 2\varphi_Z\left(p_{66}\cos^2\theta_X\sin 2\varphi_Z + p_{14}\sin 2\theta_X\cos\varphi_Z\right)
\end{aligned} \tag{7}$$

When the interaction plane is rotated around the $X$ axis, the components of the strain tensor are given by the formulas

$$\begin{aligned}
&e''_1 = e_1\cos^2\theta_X, \ e''_2 = e_1\sin^2\theta_X\sin^2\varphi_X, \ e''_3 = e_1\sin^2\theta_X\cos^2\varphi_X, \\
&e''_4 = -0.5e_1\sin^2\theta_X\sin 2\varphi_X, \ e''_5 = -0.5e_1\sin 2\theta_X\cos\varphi_X, \ e''_6 = 0.5\sin 2\theta_X\sin\varphi_X,
\end{aligned} \tag{8}$$

while the EEC becomes

$$\begin{aligned}
p_{ef}^{(I)} = &\left[\frac{\sin\varphi_X\cot(\theta_X+\theta_B)}{\sqrt{1+\sin^2\varphi_X\cot^2(\theta_X+\theta_B)}}\right]^2 \begin{pmatrix} p_{11}\cos^2\theta_X + (p_{12}\sin^2\varphi_X + p_{13}\cos^2\varphi_X) \\ + p_{14}\sin 2\varphi_X)\sin^2\theta_X \end{pmatrix} \\
&+ \left[\frac{1}{\sqrt{1+\sin^2\varphi_X\cot^2(\theta_X+\theta_B)}}\right]^2 \begin{pmatrix} p_{21}\cos^2\theta_X + (p_{22}\sin^2\varphi_X + p_{23}\cos^2\varphi_X) \\ - p_{14}\sin 2\varphi_X)\sin^2\theta_X \end{pmatrix} \\
&+ \left[\frac{2\sin\varphi_X\cot(\theta_X+\theta_B)}{1+\sin^2\varphi_X\cot^2(\theta_X+\theta_B)}\right]^2 (p_{66}\sin\varphi_X + p_{14}\cos\varphi_X)\sin 2\theta_X.
\end{aligned} \tag{9}$$

When the longitudinal AW propagates in the $YZ$ plane, the components of the strain tensor are equal to $e_2' = e_2 \cos^2 \theta_Y$, $e_3' = e_2 \sin^2 \theta_Y$ and $e_4' = -0.5 e_2 \sin 2\theta_Y$, while the EEC reads as $p_{ef}^{(I)} = p_{12} \cos^2 \theta_Y + p_{13} \sin^2 \theta_Y - p_{14} \sin 2\theta_Y$. Finally, when the interaction plane rotates around the $Y$ axis by the angle $\varphi_Y$, the components of the strain tensor caused by the AW and the EEC are as follows:

$$e_1''' = e_2 \sin^2 \theta_Y \sin^2 \varphi_Y, \; e_2''' = e_2 \cos^2 \theta_Y, \; e_3''' = e_2 \sin^2 \theta_Y \cos^2 \varphi_Y,$$
$$e_4''' = -0.5 e_2 \sin 2\theta_Y \cos \varphi_Y, \; e_5''' = -0.5 e_2 \sin^2 \theta_Y \sin 2\varphi_Y, \; e_6''' = 0.5 e_2 \sin 2\theta_Y \sin \varphi_Y, \quad (10)$$

$$p_{ef}^{(I)} = \left[ \frac{1}{\sqrt{1 + \sin^2 \varphi_Y \cot^2(\theta_Y + \theta_B)}} \right]^2 \left( \begin{array}{c} p_{12} \cos^2 \theta_Y + (p_{11} \sin^2 \varphi_Y + p_{13} \cos^2 \varphi_Y) \sin^2 \theta_Y \\ + p_{14} \sin 2\theta_Y \cos \varphi_Y \end{array} \right)$$
$$+ \left[ \frac{\sin \varphi_Y \cot(\theta_Y + \theta_B)}{\sqrt{1 + \sin^2 \varphi_Y \cot^2(\theta_Y + \theta_B)}} \right]^2 \left( \begin{array}{c} p_{11} \cos^2 \theta_Y + (p_{21} \sin^2 \varphi_Y + p_{23} \cos^2 \varphi_Y) \sin^2 \theta_Y \\ - p_{14} \sin 2\theta_Y \cos \varphi_Y \end{array} \right) \quad . \quad (11)$$
$$+ \left[ \frac{2 \sin \varphi_Y \cot(\theta_Y + \theta_B)}{1 + \sin^2 \varphi_Y \cot^2(\theta_Y + \theta_B)} \right]^2 (p_{66} \sin 2\theta_Y \sin \varphi_Y + p_{14} \sin^2 \theta_Y \sin 2\varphi_Y)$$

As a result, the AOFM for the AO interactions of the type ($I$) can be represented as

$$M_2^{(I)} = \frac{n^6 \left\{ p_{ef}^{(I)} \right\}^2}{\rho \left[ v_{QL}(\theta_{X,Y}, \varphi_{X,Y,Z}) \right]^3}, \quad (12)$$

where $v_{QL}(\theta_{X,Y}, \varphi_{X,Y,Z})$ defines the change in the AW velocity occurring in the $XZ'$, $YZ'$ and $X'Z$ planes.

Let us consider the type ($II$) of the AO interactions, when the longitudinal AW $v_{11} = v_{QL}$ propagating along the $X$ axis interacts with the optical wave, of which the electric induction vector lies in the $XZ$ plane at the angle $\theta_B$ with respect to the $X$ axis ( $D_3 = D \sin \theta_B$ and $D_1 = D \cos \theta_B$ ). Here $\theta_B$ is chosen to be equal to 0.1 deg. The electric field of the diffracted optical wave is given by

$$\begin{cases} E_1 = \Delta B_1 D_1 = p_{11} e_1 D_1 \\ E_3 = \Delta B_3 D_3 = p_{31} e_1 D_3 \end{cases} . \quad (13)$$

The strain tensor components involved are given by Eqs. (5). The EEC for the interaction plane $XZ$ reads as

$$p_{ef}^{(II)} = \cos^2 (\theta_B + \theta_X) \left[ p_{11} \cos^2 \theta_X + p_{13} \sin^2 \theta_X \right]$$
$$+ \sin^2 (\theta_B + \theta_X) \left[ p_{31} \cos^2 \theta_X + p_{33} \sin^2 \theta_X \right] + 2 p_{55} \sin \left( 2(\theta_B + \theta_X) \right) \sin \theta_X \cos \theta_X. \quad (14)$$

The relation for $p_{ef}^{(II)}$ in an arbitrary interaction plane $XZ$ reduces to

$$p_{ef}^{(II)} = \cos^2 \varphi_Z \left\{ \begin{array}{l} \cos^2(\theta_B + \theta_X)[(p_{11}\cos^2\varphi_Z + p_{12}\sin^2\varphi_Z)\cos^2\theta_X + p_{13}\sin^2\theta_X + p_{14}\sin 2\theta_X \sin\varphi_Z] \\ + \sin^2(\theta_B + \theta_X)[(p_{31}\cos^2\varphi_Z + p_{32}\sin^2\varphi_Z)\cos^2\theta_X + p_{33}\sin^2\theta_X] \\ + \sin(2(\theta_B + \theta_X))[p_{55}\sin 2\theta_X \cos\varphi_Z + p_{41}\cos^2\theta_X \sin 2\varphi_Z] \end{array} \right\}$$

$$+ \sin^2 \varphi_Z \left\{ \begin{array}{l} \cos^2(\theta_B + \theta_X)\left[(p_{21}\cos^2\varphi_Z + p_{22}\sin^2\varphi_Z)\cos^2\theta_X + p_{23}\sin^2\theta_X - p_{14}\sin 2\theta_X \sin\varphi_Z\right] \\ + \sin^2(\theta_B + \theta_X)\left[(p_{31}\cos^2\varphi_Z + p_{32}\sin^2\varphi_Z)\cos^2\theta_X + p_{33}\sin^2\theta_X\right] \\ + \sin(2(\theta_B + \theta_X))[p_{44}\sin 2\theta_X \sin\varphi_Z + p_{41}\sin 2\theta_X \cos 2\varphi_Z] \end{array} \right\}. \quad (15)$$

Similarly, the $p_{ef}^{(II)}$ parameter for arbitrary interaction planes $XZ'$ and $YZ'$ may be represented as

$$p_{ef}^{(II)} = (1 - \cos^2\varphi_X \cos^2(\theta_X + \theta_B))(p_{31}\cos^2\theta_X + p_{32}\sin^2\theta_X \sin^2\varphi_X + p_{33}\sin^2\theta_X \cos^2\varphi_X)$$
$$+ \cos^2\varphi_X \cos^2(\theta_X + \theta_B)(p_{11}\cos^2\theta_X + p_{12}\sin^2\theta_X \sin^2\varphi_X + p_{13}\sin^2\theta_X \cos^2\varphi_X + p_{14}\sin^2\theta_X \sin 2\varphi_X) \quad (16)$$
$$+ \sqrt{1 - \cos^2\varphi_X \cos^2(\theta_X + \theta_B)} \cos(\theta_X + \theta_B)\cos\varphi_X (p_{55}\sin 2\theta_X \cos\varphi_X + p_{41}\sin 2\theta_X \sin 2\varphi_X),$$

$$p_{ef}^{(II)} = (1 - \cos^2\varphi_Y \cos^2(\theta_Y + \theta_B))(p_{31}\cos^2\theta_Y + p_{32}\sin^2\theta_Y \sin^2\varphi_Y + p_{33}\sin^2\theta_Y \cos^2\varphi_Y)$$
$$+ \cos^2\varphi_Y \cos^2(\theta_Y + \theta_B)(p_{11}\cos^2\theta_Y + p_{12}\sin^2\theta_Y \sin^2\varphi_Y + p_{13}\sin^2\theta_Y \cos^2\varphi_Y - p_{14}\sin 2\theta_Y \cos\varphi_Y) \quad (17)$$
$$+ \sqrt{1 - \cos^2\varphi_Y \cos^2(\theta_Y + \theta_B)} \cos(\theta_Y + \theta_B)\cos\varphi_Y (p_{44}\sin 2\theta_Y \cos\varphi_Y + p_{41}(\sin^2\theta_Y \sin^2\varphi_Y - \cos^2\theta_Y)),$$

respectively. Thus, the AOFM for the AO interaction of the type ($II$) is given by the formula

$$M_2^{(II)} = \frac{n^6 \left\{ p_{ef}^{(II)} \right\}^2}{\rho \left[ v_{QL}(\theta_{X,Y}, \varphi_{X,Y,Z}) \right]^3}. \quad (18)$$

The AO interactions of the type ($III$) can be implemented in the LiNbO$_3$ crystals whenever the wave QT$_1$ interacts with the optical wave polarized along the $Y$ axis. After rotating the AW vector by the angle $\theta_X$, the following strain tensor components arise:

$$e_1' = e_5 \sin\theta_X \cos\theta_X, \quad e_3' = -e_5 \sin\theta_X \cos\theta_X, \quad e_5' = e_5 \cos 2\theta_X. \quad (19)$$

The EEC for the interaction in the $XZ$ plane reads as

$$p_{ef}^{(III)} = (p_{12} - p_{13})\sin\theta_X \cos\theta_X. \quad (20)$$

The EEC for the AO interaction of the transverse AW $v_{13}$ with the incident optical wave polarized parallel to the $Y$ axis in the $XZ$, $XZ'$ and $YZ'$ planes are given by

$$p_{ef}^{(III)} = \sin^2\varphi_Z \left( (p_{11}\cos^2\varphi_Z + p_{12}\sin^2\varphi_Z - p_{13})\cos\theta_X \sin\theta_X - p_{14}\cos 2\theta_X \sin\varphi_Z \right)$$
$$+ \cos^2\varphi_Z \left( (p_{21}\cos^2\varphi_Z + p_{22}\sin^2\varphi_Z - p_{23})\cos\theta_X \sin\theta_X + p_{14}\cos 2\theta_X \sin\varphi_Z \right) \quad (21)$$
$$+ \sin 2\varphi_Z (p_{14}\cos 2\theta_X \cos\varphi_Z - 0.5 p_{66}\sin 2\theta_X \sin 2\varphi_Z),$$

$$p_{ef}^{(III)} = \left[\frac{\sin\varphi_X \cot(\theta_X + \theta_B)}{\sqrt{1 + \sin^2\varphi_X \cot^2(\theta_X + \theta_B)}}\right]^2 \left(\begin{array}{c} 0.5(p_{11} - p_{12}\sin^2\varphi_X - p_{13}\cos^2\varphi_X)\sin 2\theta_X \\ -0.5p_{14}\sin 2\theta_X \sin 2\varphi_X \end{array}\right)$$

$$+ \left[\frac{1}{\sqrt{1 + \sin^2\varphi_X \cot^2(\theta_X + \theta_B)}}\right]^2 \left(\begin{array}{c} 0.5(p_{21} - p_{22}\sin^2\varphi_X - p_{23}\cos^2\varphi_X)\sin 2\theta_X \\ +0.5p_{14}\sin 2\theta_X \sin 2\varphi_X \end{array}\right) \quad (22)$$

$$+ \left[\frac{2\sin\varphi_X \cot(\theta_X + \theta_B)}{1 + \sin^2\varphi_X \cot^2(\theta_X + \theta_B)}\right]^2 (p_{66}\sin\varphi_X + p_{14}\cos\varphi_X)\cos 2\theta_X,$$

and

$$p_{ef}^{(III)} = \left[\frac{\sin\varphi_Y \cot(\theta_Y + \theta_B)}{\sqrt{1 + \sin^2\varphi_Y \cot^2(\theta_Y + \theta_B)}}\right]^2 \left(\begin{array}{c} 0.5(p_{11} - p_{12}\sin^2\varphi_Y - p_{13}\cos^2\varphi_Y)\sin 2\theta_Y \\ +p_{14}\cos 2\theta_Y \cos\varphi_Y \end{array}\right)$$

$$+ \left[\frac{1}{\sqrt{1 + \sin^2\varphi_Y \cot^2(\theta_Y + \theta_B)}}\right]^2 \left(\begin{array}{c} 0.5(-p_{11}\sin^2\varphi_Y + p_{12} - p_{13}\cos^2\varphi_Y)\sin 2\theta_Y \\ -p_{14}\cos 2\theta_Y \cos\varphi_Y \end{array}\right) \quad (23)$$

$$+ \left[\frac{2\sin\varphi_Y \cot(\theta_Y + \theta_B)}{1 + \sin^2\varphi_Y \cot^2(\theta_Y + \theta_B)}\right]^2 (p_{66}\cos 2\theta_Y \sin\varphi_Y - 0.5p_{14}\sin 2\theta_Y \sin 2\varphi_Y),$$

respectively. As a result, the AOFM becomes as follows:

$$M_2^{(III)} = \frac{n^6 \left\{p_{ef}^{(IV)}\right\}^2}{\rho \left[v_{QT_i}(\theta_{X,Y}, \varphi_{X,Y,Z})\right]^3}. \quad (24)$$

Now we consider the interactions of the type (*IV*). The components of the strain tensor and the EEC corresponding to the AW propagation and AO interaction occurring in the *XZ* plane are as follows:

$$e_1' = \sin 2\theta_X e_5, \ e_3' = -\sin 2\theta_X e_5, \ e_5' = \cos 2\theta_X e_5, \ p_{ef}^{(IV)} = 0.5(p_{11} - p_{13})\sin 2\theta_X. \quad (25)$$

When the interaction plane rotates around the *Z* axis, the following components of the strain tensor appear:

$$e_1' = 0.5\sin 2\theta_X \cos^2\varphi_Z e_5, \ e_2' = 0.5\sin 2\theta_X \sin^2\varphi_Z e_5, \ e_3' = -0.5\sin 2\theta_X e_5,$$
$$e_4' = \cos 2\theta_X \sin\varphi_Z e_5, \ e_5' = \cos 2\theta_X \cos\varphi_Z e_5, \ e_6' = 0.5\sin 2\theta_X \sin 2\varphi_Z e_5. \quad (26)$$

The corresponding EEC is given by

$$
p_{ef}^{(IV)} = \cos^2 \varphi_Z \begin{bmatrix} \cos^2(\theta_B + \theta_X)[(p_{11}\cos^2\varphi_Z + p_{12}\sin^2\varphi_Z - p_{13})\sin\theta_X\cos\theta_X - p_{14}\cos2\theta_X\sin\varphi_Z] \\ + \sin^2(\theta_B + \theta_X)[p_{31}\cos^2\varphi_Z + p_{32}\sin^2\varphi_Z - p_{33}]\sin\theta_X\cos\theta_X \\ + \sin(2(\theta_B + \theta_X))[p_{55}\cos2\theta_X\cos\varphi_Z - 0.5p_{41}\sin2\theta_X\sin2\varphi_Z)] \end{bmatrix}
$$
$$
+ \sin^2\varphi_Z \begin{bmatrix} \cos^2(\theta_B + \theta_X)[(p_{21}\cos^2\varphi_Z + p_{22}\sin^2\varphi_Z - p_{23})\sin\theta_X\cos\theta_X + p_{14}\cos2\Theta\sin\varphi_Z] \\ + \sin^2(\theta_B + \theta_X)[p_{31}\cos^2\varphi_Z + p_{32}\sin^2\varphi_Z - p_{33}]\sin\theta_X\cos\theta_X \\ + \sin(2(\theta_B + \theta_X))(-p_{44}\cos2\theta_X\sin\varphi_Z + p_{41}\sin2\theta_X\cos2\varphi_Z) \end{bmatrix}, \quad (27)
$$

whereas Eq. (21) for arbitrary $XZ'$ and $YZ'$ planes is replaced by

$$
p_{ef}^{(IV)} = (1 - \cos^2\varphi_X\cos^2(\theta_X + \theta_B))(p_{31} - p_{32}\sin^2\varphi_X - p_{33}\cos^2\varphi_X)\sin\theta_X\cos\theta_X
$$
$$
+ \cos^2\varphi_X\cos^2(\theta_X + \theta_B)((p_{11} - p_{12}\sin^2\varphi_X - p_{13}\cos^2\varphi_X)\sin\theta_X\cos\theta_X - 0.5p_{14}\sin2\theta_X\sin2\varphi_X) \quad (28)
$$
$$
+ \sqrt{(1 - \cos^2\varphi_X\cos^2(\theta_X + \theta_B))}\cos(\theta_X + \theta_B)^2\cos\varphi_X\cos2\theta_X(p_{55}\cos\varphi_X + p_{41}\sin\varphi_X),
$$

$$
p_{ef}^{(IV)} = (1 - \cos^2\varphi_Y\cos^2(\theta_Y + \theta_B))(-p_{31} + p_{32}\sin^2\varphi_Y - p_{33}\cos^2\varphi_Y)\sin\theta_Y\cos\theta_Y
$$
$$
+ \cos^2\varphi_Y\cos^2(\theta_Y + \theta_B)((-p_{21} + p_{22}\sin^2\varphi_Y - p_{23}\cos^2\varphi_Y)\sin\theta_Y\cos\theta_Y - p_{14}\cos2\theta_Y\cos\varphi_Y) + \quad (29)
$$
$$
+ \sqrt{(1 - \cos^2\varphi_Y\cos^2(\theta_Y + \theta_B))}\cos(\theta_Y + \theta_B)^2\cos\varphi_Y(p_{44}\cos2\theta_Y\cos\varphi_Y - 0.5p_{41}\sin2\theta_Y(1 + \sin^2\varphi_Y)),
$$

respectively. Then the AOFM reads as

$$
M_2^{(IV)} = \frac{n^6 \left\{ p_{ef}^{(IV)} \right\}^2}{\rho \left[ v_{QT_1}(\theta_{X,Y}, \varphi_{X,Y,Z}) \right]^3}. \quad (30)
$$

Let us proceed to the AO interaction of optical waves with the AW $QT_2$ ($v_{12} = v_{QT_2}$) propagating along the $X$ axis and polarized along the $Y$ axis. This corresponds to the AO interactions of the types ($V$ and $VI$), and the AW produces the strain tensor component $e_6$. In general, depending on the orientation of the electric induction of the incident optical wave (i.e., availability of the components $D_2$, $D_3 = D\sin\theta_B$ and $D_1 = D\cos\theta_B$), the AO interactions of the types ($V$) or ($VI$) are dealt with. For the case of interaction of the type ($V$), the strain tensor includes the two components dependent upon the AW vector orientation:

$$
e_6' = e_6\cos\theta_X, \quad e_4' = -e_6\sin\theta_X. \quad (31)
$$

Then the increment of the optical-frequency impermeability tensor reduces respectively to

$$
\Delta B_2 = -p_{24}e_6\sin\theta_X \quad (32)
$$

and

$$
p_{ef}^{(V)} = -p_{24}\sin\theta_X = p_{14}\sin\theta_X. \quad (33)
$$

Rotation of the interaction plane around the $Z$ axis results in the corrections

$$\Delta B_2' = -\cos 2\varphi_Z \left( p_{66} \cos \theta_X \sin 2\varphi_Z - p_{14} \sin \theta_X \cos \varphi_Z \right)$$
$$+ \sin 2\varphi_Z \left( p_{66} \cos \theta_X \cos 2\varphi_Z - p_{14} \sin \theta_X \cos \varphi_Z \right) \quad (34)$$

$$\Delta n = -\frac{1}{2} n^3 \begin{bmatrix} -\cos 2\varphi_Z \left( p_{66} \cos \theta_X \sin 2\varphi_Z - p_{14} \sin \theta_X \cos \varphi_Z \right) \\ + \sin 2\varphi_Z \left( p_{66} \cos \theta_X \cos 2\varphi_Z - p_{14} \sin \theta_X \cos \varphi_Z \right) \end{bmatrix}, \quad (35)$$

with the EEC being equal to

$$p_{ef}^{(V)} = -\cos 2\varphi_Z \left( p_{66} \cos \theta_X \sin 2\varphi_Z - p_{14} \sin \theta_X \cos \varphi_Z \right)$$
$$+ \sin 2\varphi_Z \left( p_{66} \cos \theta_X \cos 2\varphi_Z - p_{14} \sin \theta_X \cos \varphi_Z \right). \quad (36)$$

Finally, rotation of the interaction plane around the $X$ and $Y$ axis results in

$$p_{ef}^{(V)} = \left[ \frac{\sin \varphi_X \cot(\theta_X + \theta_B)}{\sqrt{1 + \sin^2 \varphi_X \cot^2(\theta_X + \theta_B)}} \right]^2 \left( (p_{12} - p_{13}) \sin \theta_X \sin \varphi_X \cos \varphi_X - p_{14} \sin \theta_X \cos 2\varphi_X \right)$$
$$+ \left[ \frac{1}{\sqrt{1 + \sin^2 \varphi_X \cot^2(\theta_X + \theta_B)}} \right]^2 \left( (p_{11} - p_{13}) \sin \theta_X \sin \varphi_X \cos \varphi_X + p_{14} \sin \theta_X \cos 2\varphi_X \right) \quad (37)$$
$$+ \left[ \frac{2 \sin \varphi_X \cot(\theta_X + \theta_B)}{1 + \sin^2 \varphi_X \cot^2(\theta_X + \theta_B)} \right] (p_{66} \cos \varphi_X + p_{14} \sin \varphi_X) \cos \theta_X$$

and

$$p_{ef}^{(V)} = \left[ \frac{1}{\sqrt{1 + \sin^2 \varphi_Y \cot^2(\theta_Y + \theta_B)}} \right]^2 \left( (p_{11} - p_{13}) \sin \theta_Y \sin \varphi_Y \cos \varphi_Y - p_{14} \cos \theta_Y \sin \varphi_Y \right)$$
$$+ \left[ \frac{\sin \varphi_Y \cot(\theta_Y + \theta_B)}{\sqrt{1 + \sin^2 \varphi_Y \cot^2(\theta_Y + \theta_B)}} \right]^2 \left( (p_{12} - p_{13}) \sin \theta_Y \sin \varphi_Y \cos \varphi_Y + p_{14} \cos \theta_Y \sin \varphi_Y \right) \quad (38)$$
$$+ \left[ \frac{2 \sin \varphi_Y \cot(\theta_Y + \theta_B)}{1 + \sin^2 \varphi_Y \cot^2(\theta_Y + \theta_B)} \right] (p_{66} \cos \theta_Y \cos \varphi_Y + p_{14} \sin \theta_Y \cos 2\varphi_Y),$$

respectively. Then the AOFM is as follows:

$$M_2^{(V)} = \frac{n^6 \left\{ p_{ef}^{(V)} \right\}^2}{\rho \left[ v_{QT_2}(\theta_{X,Y}, \varphi_{X,Y,Z}) \right]^3}. \quad (39)$$

Our final step is to consider the type ($VI$) of AO interactions, when the polarizations of the incident and diffracted optical waves belong to the $XZ$ plane. According to Eqs. (31), the EEC at the rotation of interaction plane around $Z$ axis is given by

$$p_{ef}^{(VT)} = \cos^2 \varphi_Z [\cos^2(\theta_B + \theta_X)((p_{11} - p_{12})\cos\theta_X \cos\varphi_Z \sin\varphi_Z - p_{14}\sin\theta_X \cos\varphi_Z)$$
$$+ \sin 2(\theta_B + \theta_X)(-p_{44}\sin\theta_X \sin\varphi_Z + p_{41}\cos\theta_X \cos 2\varphi_Z)]$$
$$+ \sin^2 \varphi_Z [\cos^2(\theta_B + \theta_X)((p_{12} - p_{11})\cos\theta_X \cos\varphi_Z \sin\varphi_Z + p_{14}\sin\theta_X \cos\varphi_Z)$$
$$+ \sin 2(\theta_B + \theta_X)(-p_{44}\sin\theta_X \cos\varphi_Z + p_{41}\cos\theta_X \sin 2\varphi_Z)] + \sin 2\varphi_Z (p_{66}\cos\theta_X \cos 2\varphi_Z - p_{14}\sin\theta_X \sin\varphi_Z)$$
$$. \quad (40)$$

Rotation of the interaction plane around the $X$ axis results in

$$p_{ef}^{(VT)} = (1 - \cos^2\varphi_X \cos^2(\theta_X + \theta_B))(p_{32} - p_{33})\sin\theta_X \sin\varphi_X \cos\varphi_X$$
$$+ \cos^2\varphi_X \cos^2(\theta_X + \theta_B)((p_{12} - p_{13})\sin\theta_X \sin\varphi_X \cos\varphi_X - p_{14}\sin\theta_X \cos\varphi_X) \quad , \quad (41)$$
$$+ \sqrt{1 - \cos^2\varphi_X \cos^2(\theta_X + \theta_B)}\cos\varphi_X \cos(\theta_X + \theta_B)(p_{55}\sin\varphi_X + p_{41}\cos\varphi_X)\cos\theta_X$$

whereas rotation of the interaction plane $YZ$ around the $Y$ axis yields

$$p_{ef}^{(VT)} = (1 - \cos^2\varphi_Y \cos^2(\theta_Y + \theta_B))(p_{31} - p_{33})\sin\theta_Y \sin\varphi_Y \cos\varphi_Y$$
$$+ \cos^2\varphi_Y \cos^2(\theta_Y + \theta_B)((p_{21} - p_{23})\sin\theta_Y \sin\varphi_Y \cos\varphi_Y + p_{14}\cos\theta_Y \sin\varphi_Y) \quad . \quad (42)$$
$$+ 2\sqrt{1 - \cos^2\varphi_Y \cos^2(\theta_Y + \theta_B)}\cos\varphi_Y \cos(\theta_Y + \theta_B)(-p_{44}\sin\varphi_Y \cos\theta_Y + 0.5 p_{41}\sin 2\varphi_Y \sin\theta_Y)$$

Then the AOFM becomes as follows:

$$M_2^{(VT)} = \frac{n^6 \left\{ p_{ef}^{(VT)} \right\}^2}{\rho \left[ v_{QT_2}(\theta_{X,Y}, \varphi_{X,Y,Z}) \right]^3} \quad . \quad (43)$$

## 3.2. Results of analysis

To demonstrate the impact of anisotropy on the AOFM value, below we present a number of examples for the maximum AOFM values, which can be reached for the cases of all of the six different types of isotropic AO interactions. The dependences of the AOFM on the direction of AW propagation calculated at different orientations of the interaction planes are displayed in Fig. 2.

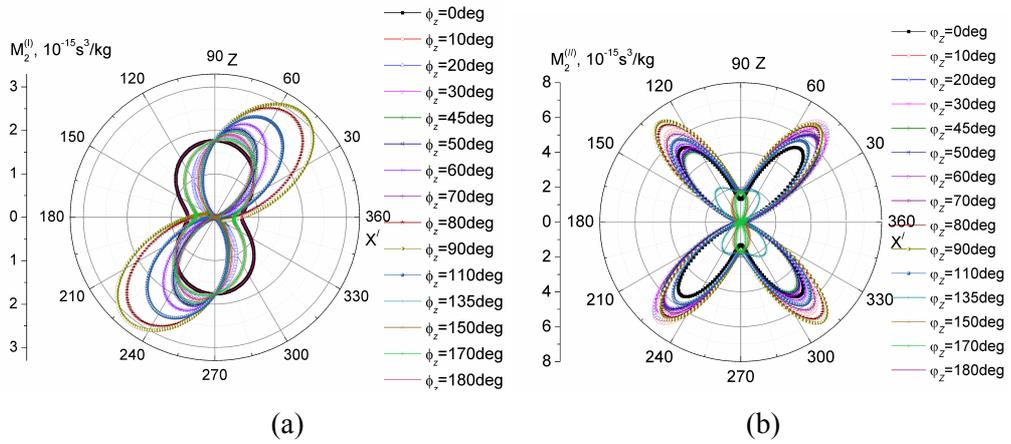

(a)                                        (b)

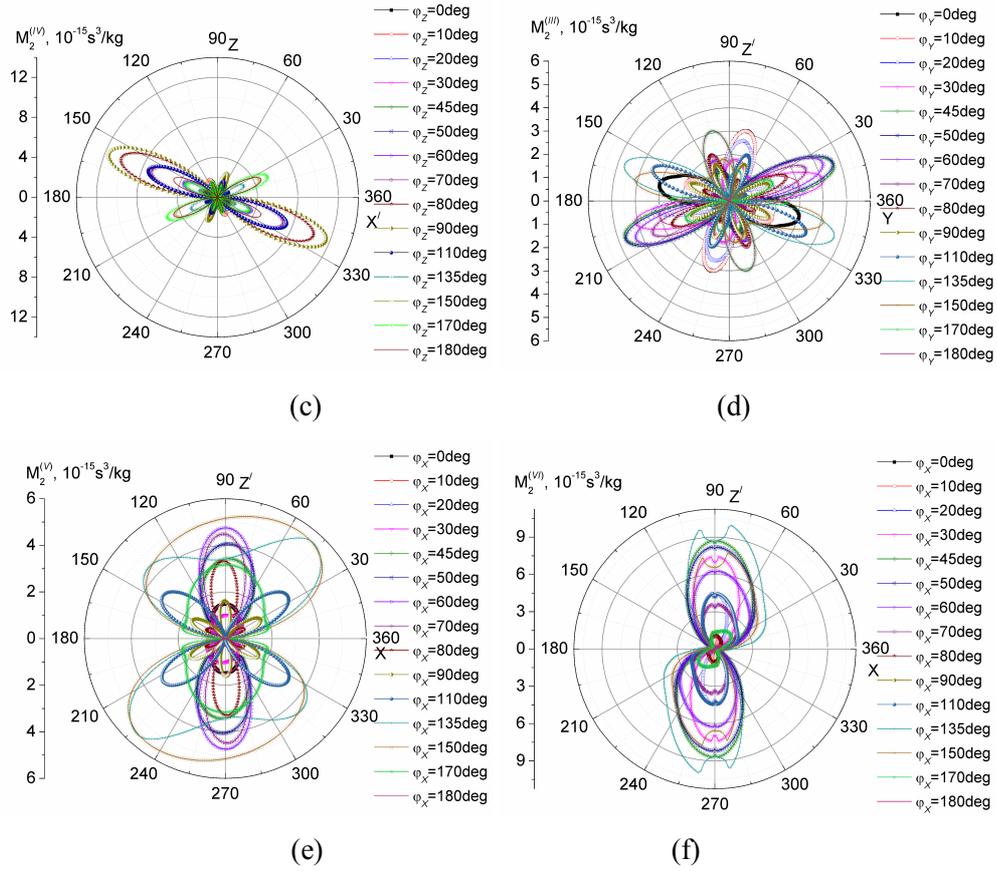

Fig. 2. Dependencies of AOFM on the $\theta_X$ (or $\theta_Y$) angle calculated at different orientations of the interaction planes, which are determined by $\varphi_X$, $\varphi_Y$ and $\varphi_Z$ angles: (a) type (*I*); (b) type (*II*); (c) type (*III*); (d) type (*IV*); (e) type (*V*), and (f) type (*VI*) of AO interactions.

The maximum AOFM value for the type (*I*) of AO interactions is equal to $3.11\times10^{-15}\,\mathrm{s}^3/\mathrm{kg}$. This value is achieved in the case of AO interaction with the longitudinal AW propagating in the *YZ* plane at the angle of 51 deg with respect to the *Y* axis ($\varphi_Z = 90$ deg – see Fig. 2a and Table 1). The maximum AOFM for the interaction type (*II*), $7.36\times10^{-15}\,\mathrm{s}^3/\mathrm{kg}$, is achieved at $\varphi_Z = 10$ deg and $\theta_X = 50$, 230 deg (see Fig. 2b and Table 1). We obtain almost the same AOFM ($7.34\times10^{-15}\,\mathrm{s}^3/\mathrm{kg}$) when the geometry is given by $\varphi_Z = 90$ deg and $\theta_X = 130$, 310 deg. The AO interaction of the type (*III*) is characterized by the maximum equal to $4.72\times10^{-15}\,\mathrm{s}^3/\mathrm{kg}$. It corresponds to $\varphi_Y = 45$, 135 deg and $\theta_Y = 19$, 199, 161, 341 deg. For the type (*IV*) of AO interactions, the maximum AOFM is equal to $11.62\times10^{-15}\,\mathrm{s}^3/\mathrm{kg}$ (see

Fig. 2d). The appropriate angles describing the experimental geometry are collected in Table 1.

Table 1. Maximum AOFM values for the LiNbO$_3$ crystals calculated for the six types of isotropic AO interactions listed in Section 2.

| Type of AO interaction | (I) | (II) | (III) | (IV) | (V) | (VI) |
|---|---|---|---|---|---|---|
| $M_2$, $10^{-15}$ s$^3$/kg | 3.11 | 7.36 | 4.72 | 11.62 | 5.58 | 10.03 |
| $\theta_{X,Y}$, deg | $\theta_X = 51$ | $\theta_X = 50, 230$ | $\theta_Y = 19, 199, 161, 341$ | $\theta_X = 157.5, 337.5$ | $\theta_X = 55, 235$ | $\theta_X = 82.5, 262.5$ |
| $\varphi_{X,Y,Z}$, deg | $\varphi_Z = 90$ | $\varphi_Z = 10$ | $\varphi_Y = 45, 135$ | $\varphi_Z = 90$ | $\varphi_X = 150$ | $\varphi_X = 135$ |

The AOFM, $4.69 \times 10^{-15}$ s$^3$/kg, has to be observed at $\varphi_Y = 50$ deg and $\theta_Y = 21, 201$ deg. For the type (V) of AO interactions, the maximum value of AOFM ($5.58 \times 10^{-15}$ s$^3$/kg) is peculiar for $\varphi_X = 150$ deg and $\theta_X = 55, 235$ deg. Finally, for the type (VI) we have the highest AOFM, $10.03 \times 10^{-15}$ s$^3$/kg, under the conditions $\theta_X = 82.5, 262.5$ deg and $\varphi_X = 135$ deg. Hence, the highest AOFM value achievable under the conditions of isotropic AO diffraction occurs for the interactions of the type (IV), being equal to $11.62 \times 10^{-15}$ s$^3$/kg. On the other hand, the type (VI) of AO diffraction is also characterized by high enough AOFM.

Now let us analyze specific reasons why these AOFM values are so high (see Eq. (1)). As seen from Fig. 3b, the highest AOFM achieved for the case of AO interactions of the type (IV) is determined solely by the extremum of the EEC. Indeed, the cube of the AW slowness becomes the smallest at $\theta_X = 157.5$ and $337.5$ deg (see Fig. 3a). On the other hand, the high AOFM value observed for the type (VI) of AO interactions (see Fig. 3c, d) represents a combined effect of the extreme AW slowness and the high EEC value, which both occur at $\theta_X = 82.5$ and $262.5$ deg. As seen from Fig. 3a, c, the dependences of the AW slowness on the polar angle are not monotonous. This fact is caused by a nonlinear trigonometric relationship between the polar angle and the orientation angle of the AW vector. This situation typically takes place if one considers the diffraction processes that involve elastooptic disturbance of the extraordinary refractive index (see Ref. [18] for more details).

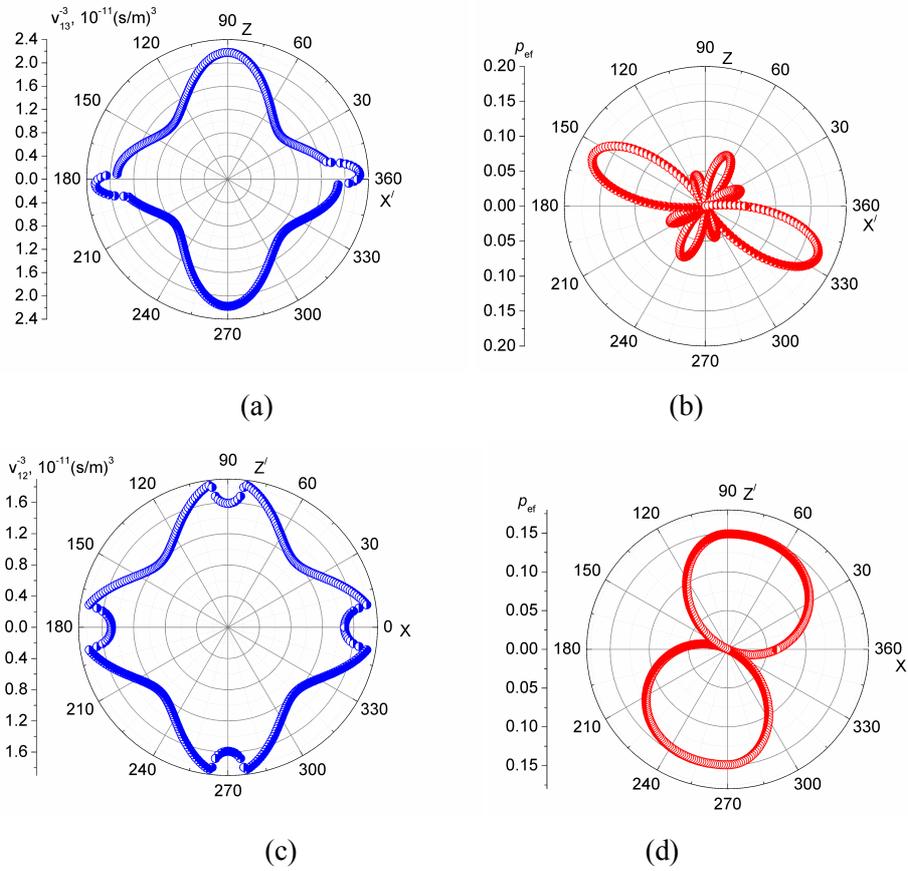

(a)

(b)

(c)

(d)

Fig. 3. Dependences of cube of the AW slowness (a, c) and EEC (b, d) on the AW propagation directions given by the angle $\theta_X$: (a, b) $QT_1$ and (c, d) $QT_2$. Panels (a, b) correspond to $\varphi_Z = 90$ deg and panels (c, d) to $\varphi_X = 135$ deg

To test the results of our analysis, we make a comparison with the available experimental data (see Refs. [13, 26]). As seen from Table 2, some of our results agree well with the experimental data (items #1–3). However, the experimental geometries given by items #4, 5 and 6 contradict the experimental results obtained using the Dixon–Cohen method. In the geometry #4, the elastooptic coefficient $p_{33} = 0.141 \pm 0.013$ obtained by us [21] with the interferometric technique is almost twice as large as the $p_{33}$ reported in Refs. [26, 27] basing on the Dixon–Cohen method. On the other hand, the same coefficient obtained in Ref. [28] is equal to $0.118 \pm 0.020$, which agrees with our data within the confidence intervals. For the geometry #5, the EEC is equal to $p_{13} = 0.147 \pm 0.019$, according to Ref. [21], while the au-thors of Ref. [26] have reported it to be 0.106. Finally, the EEC value obtained in Ref. [26] for the geometry #6 ($p_{11} = 0.045$) is two times larger than that reported in Ref. [21] ($p_{11} = 0.023 \pm 0.017$) and in the works [28] ($p_{11} = 0.021 \pm 0.018$) and [29] ($p_{11} = -0.026$). As a consequence, in all the cases the disagreement of our calculation results with the experimental data should be caused by the values of EECs used in Ref. [21], which seem to be more reliable than those presented in Refs. [13, 26].

Table 2. AOFM values obtained experimentally in Ref. [13, 26] for different types of isotropic AO interactions in LiNbO$_3$ crystals, along with the results of our analysis.

| # | AW | Optical wave polarization | $M_2$, $10^{-15}$ s$^3$/kg [13] | $M_2$, $10^{-15}$ s$^3$/kg [25] | $p_{ef}$ [26] | $M_2$, $10^{-15}$ s$^3$/kg (type of AO interaction), according to our data | $p_{ef}$ according to our data [21] |
|---|---|---|---|---|---|---|---|
| 1 | $v_{11}$ | $\parallel Z$ | 2.74 | 2.05 | 0.138 | 2.09 (II) | $p_{31} = 0.157$ |
| 2 | $v_{11}$ | $\parallel Y$ | 0.56 | 0.783 | 0.096 | 0.614 (I) | $p_{21} = 0.076$ |
| 3 | $v_{22}$ | $\parallel Z$ | | 2.03 | | 2.09 (II) | $p_{31} = 0.157$ |
| 4 | $v_{33}$ | $\parallel Z$ | 0.46 | 0.466 | 0.076 | 1.3 (II) | $p_{33} = 0.141$ |
| 5 | $v_{33}$ | $\perp Z$ | 0.64 | 0.725 | 0.106 | 1.76 (I) | $p_{13} = 0.147$ |
| 6 | $v_{11}$ | $\parallel X$ | 0.14 | 0.172 | 0.045 | 0.045 (II) | $p_{11} = -0.023$ |

Finally, we are to stress that the maximum AOFM value typical for the isotropic AO interactions in the LiNbO$_3$ crystals is equal to $11.62 \times 10^{-15}$ s$^3$/kg rather than $10.4 \times 10^{-15}$ s$^3$/kg as stated in Ref. [15]. In addition we have to notice that the value of AOFM, which is equal to $10.4 \times 10^{-15}$ s$^3$/kg, was obtained in [15] with ignoring of the momentum conservation law. Namely, the acoustic wave vector does not belong to the plane of the incident and diffracted wave vectors of the optical waves. More over, in our notation this value correspond to our V type of AO interaction and as it can be seen (Table 1), the maximum value of AOFM that corresponds to this type of interaction is equal to $5.58 \times 10^{-15}$ s$^3$/kg. Notice that the direction of propagation of the waves in the case of V type of interaction which is characterized by maximum value of AOFM (Table 1) completely differ from directions of wave vectors, presented in [15]. The transverse AW corresponding to the mentioned case of AO interactions propagates in the $YZ$ plane at the angle of 157.5 deg (or 337.5 deg) with respect to the $Y$ axis and is polarized in the same plane. Its velocity is equal to 3994 m/s. This is not the same AW that has been reported in Ref. [15].

# 4. Conclusions

In the present work we have developed the approach, which has been successfully applied when analyzing the AOFM anisotropy in the lithium niobate crystals for the case of isotropic AO diffraction. The working relationships obtained by us for the EEC and the AOFM can be used for the other crystals belonging to the point symmetry groups 32 and $\overline{3}$m, since the structures of the elastooptic and elastic stiffness tensors for these groups are the same as those typical for the symmetry group 3m.

We have found that the maximum AOFM value for the LiNbO$_3$ crystals is $11.62 \times 10^{-15}$ s$^3$/kg. It takes place under the conditions of isotropic AO diffraction in the following geometry of AO interactions: the shear AW propagates, with the velocity 3994 m/s, in the $YZ$ plane at the angle 157.5 deg (or 337.5 deg) with respect to the $Y$ axis, whereas the incident optical wave propagates almost normally to the AW in the $YZ$ plane, with the polarization lying in the same plane. This corresponds to the AO interactions of the type ($IV$), according to our classification. The type ($VI$) of AO interactions is also characterized by a high enough AOFM. It equals to $10.03 \times 10^{-15}$ s$^3$/kg and is reached at the conditions given by the angles $\theta_X = 82.5$, 262.5 deg and $\varphi_X = 135$ deg. In this case the shear AW has the velocity 3790 m/s, and both the optical wave and the AW are polarized in the interaction plane.

We have demonstrated that the maximum AOFM values in the LiNbO$_3$ crystals are mainly achieved owing to notable anisotropy and relatively high values of the EEC, since the AW slowness is very low. The calculated parameters obtained for the AOFM anisotropy in the LiNbO$_3$ crystals have been compared with the available experimental data and shown to agree fairly well with the latter.

The analysis of the anisotropic diffraction types in the LiNbO$_3$ crystals will be presented in the forthcoming second part of this study.

# Acknowledgement

The authors acknowledge financial support of the present work from the Ministry of Education and Science of Ukraine (the project #0115U000095).